# Experimental Demonstration of Nonlinear Frequency Division Multiplexed Transmission

Vahid Aref, Henning Bülow, Karsten Schuh, Wilfried Idler

Bell Laboratories, Alcatel-Lucent, Stuttgart, Germany, Email: firstname.lastname@alcatel-lucent.com

**Abstract** *We experimentally demonstrate an NFDM optical system with modulation over nonlinear discrete spectrum. Particularly, each symbol carries 4-bits from multiplexing two eigenvalues modulated by QPSK constellation. We show a low error performance using NFT detection with 4Gbps rate over 640km.*

## Introduction

Signal propagation in an optical fiber is usually modelled by nonlinear Schrödinger equation (NLSE) which characterizes the interplay between dispersion and Kerr nonlinearity in the channel. A well-known method to study NLSE is so-called nonlinear Fourier transform (NFT), describing the optical fields by a nonlinear spectrum[1] and resulting in a simple linear channel transfer function even for strongly nonlinear propagation along a fiber link.

The nonlinear spectrum is divided in two parts: the continuous spectrum, corresponding to the "dispersive" signal components, and the discrete spectrum, corresponding to the "solitonic" (non-dispersive) signal components. The discrete spectrum is characterized by $(\lambda_i, Q_d(\lambda_i))$ where $\lambda_i$ is an isolated spectral parameter in upper complex plane (the eigenvalue of Lax operator[3]), and $Q_d(\lambda_i)$ is the discrete spectral function at the spectral parameter $\lambda_i$.

Exploiting nonlinearity in optical systems returns two decades back when on-off keying soliton transmission was proposed[2]. Following the advances of coherent technology in the last decade, the demand for higher transmission rates attracts more interests in using nonlinear effects in communication design. In a series of papers, Yousefi and Kschischang study some general principles for modulation over nonlinear spectrum including nonlinear frequency division multiplexing (NFDM)[3-5]. NFDM transmission based on on-off keying is proposed[5] and recently verified experimentally by using 3 or 4 subcarriers (eigenvalues) and $0.5GBd$ transmission rate[6]. Simultaneously, NFDM is also investigated by Prilepsky, Turitsyn *et al.*[7] and modulation schemes over continuous spectrum are studied by Le *et al.*[8]. Recently, it is studied experimentally how to detect transmitted waveforms based on NFT in two steps: first, based on eigenvalue patterns and then, based on spectral functions[9,10].

In this paper, we experimentally demonstrate NFDM transmission based on modulation of discrete spectrum. The constellation set is generated from multiplexing two eigenvalues, $\lambda_1 = 0.6j$ and $\lambda_2 = 0.3j$. The spectral function of each eigenvalue forms a QPSK constellation,

$$Q_d(\lambda_1) = \exp\left(jk_1\frac{\pi}{2}\right), Q_d(\lambda_2) = \exp\left(j(k_2\frac{\pi}{2} + \frac{\pi}{4})\right),$$

for $k_1, k_2 = 0,1,2,3$, in total, 16 2-soliton symbols. We show that information bits can be detected with a small error rate with $4Gbps$ transmission rate over $640km$ fiber link.

## Experimental Setup

In our experiment, each 4 bits are mapped to a 2-soliton symbol of the constellation set. More precisely, each 2 bits are Grey mapped to one of the spectral functions $Q_d(\lambda_i)$ and then, the corresponding soliton is generated (offline) by an INFT algorithm, e.g. Darboux Transform[3], and stored each in 64 samples. Four of these solitons are depicted in Fig. 2(a). The others are realized from $k\frac{\pi}{2}$ phase shifts of these symbols.

The experimental setup is shown in Fig. 1. Following a $64GSa/s$ digital-to-analog converter (DAC), a drive signal is provided for a Mach-Zehnder IQ modulator which transmits a single polarization $1\ GBd$ stream into a link of up to 8 spans ($L_{\text{span}} = 80\ km$) of NZ-DSF fiber with a mean launch power of ~0dBm for each span. The received OSNR is estimated around 25dB. A signal of $2^8$ solitons is repetitively transmitted. The corresponding $2^{10}$ bit sequence was taken from a PRBS $2^{11} - 1$ bit sequence.

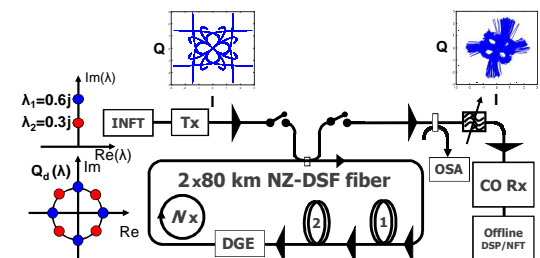

**Fig. 1:** Experimental setup: scatter plots of constellation symbols are shown at transmitter and receiver (after phase recovery) after 640 km transmission.

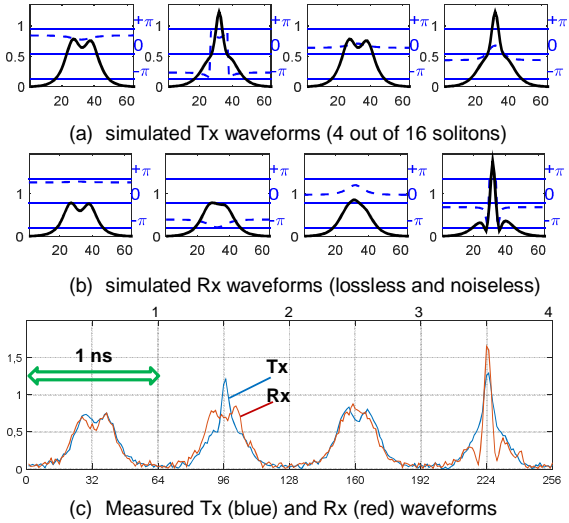

**Fig. 2: (a)** Four transmitted 2-Solitons: Solid lines show the absolute value and dashed lines show the phase. **(b)** The received solitons after 640 km over a lossless fiber. **(c)** The measured signals: back to back measurements (blue curves) and after 640 km NZ-DSF (red curves).

To simplify the carrier recovery, we use homodyne detection with a low phase noise fiber laser ($1\,kHz$ linewidth). The transmitted signal is coherently detected and sampled by a scope with $80\,Gsa/s$ sampling rate and a bandwidth of $33\,GHz$, which is followed by an offline DSP and finally an NFT algorithm to recover both QPSK spectral phases.

**Detection Performance**

In the absence of noise, NFT provides certain predictions how the discrete spectrum transforms in a lossless optical link: the eigenvalues remain the same and the discrete spectrum transforms linearly by

$$Q_d(\lambda_i; L) = Q_d(\lambda_i) \exp(-j4\lambda_i^2 L/L_0), \quad (1)$$

where $L$ is the link length and $L_0$ is the normalization factor specified by fiber physical parameters and the transmission rate $1\,GBd$. For an NZ-DSF fiber, the typical values of $\beta_2 = -5.75\,ps^2 km^{-1}$ and $\gamma = 1.6\,mW^{-1}km^{-1}$ determine $L_0$ and the required launch power $P_{\text{ideal}}$. Fig. 2(a) and (b) show four of the solitons at transmitter and at receiver after 640 km.

In practice, the optical link is neither lossless nor noiseless. Although the signals are attenuated with $\alpha \approx 0.2\,dBkm^{-1}$ in NZ-DSF fiber, we still observe a good agreement between the received signals in the experiment (see Fig. 2(c)) and those in the ideal lossless channel (see Fig. 2(a) and (b)), if the launch power is tuned about 5.7 dB beyond $P_{\text{ideal}}$ of the lossless simulations.

After data-directed frequency offset and phase offset corrections, we apply NFT to compute the spectrum of received signals. Let $(\tilde{\lambda}_i, \tilde{Q}_d(\tilde{\lambda}_i))$ denote the discrete spectrum of a received signal. We transform $\tilde{Q}_d(\tilde{\lambda}_i)$ back to the transmitter reference by Eq. (1) (by choosing $\tilde{\lambda}_i = \lambda_i$). Fig. 3 illustrates the discrete spectrum of more than $2^{16}$ received signals. The 16 plots in Fig. 3(a), labelled from 1 to 16, show the estimation of two eigenvalues $\tilde{\lambda}_1$ and $\tilde{\lambda}_2$. The first four plots correspond to solitons in Fig. 2. The blue (top) and red (bottom) dots show the deviation of $\tilde{\lambda}_1$ and $\tilde{\lambda}_2$, respectively, from $\lambda_1 = 0.6j$ and $\lambda_2 = 0.3j$ in the complex plane. Fig. 3(b) and (c) illustrate $\tilde{Q}_d(\tilde{\lambda}_1)$ and $\tilde{Q}_d(\tilde{\lambda}_2)$ corresponding to each plot in Fig. 3(a). As our detection algorithm here works only based on spectral phases, we normalize $\tilde{Q}_d(\tilde{\lambda}_i)$ to 1 to have visually clear plots.

We estimate the sent bits by the following simple algorithm. For each symbol, $\tilde{Q}_d(\tilde{\lambda}_1)$ and $\tilde{Q}_d(\tilde{\lambda}_2)$ are decoded separately according to their phase by mapping them to the closest constellation point. The decision regions are partitioned by black solid lines in polar plots in Fig. 3(b) and (c). From transmission of more than $2^{20}$ bits, we observe an average error probability less than 0.002. 85% of errors occur on symbol 10, 8% on symbol 7, and the rest only on symbol 2, 8 and 13. Note that Fig. 3 shows a strong correlation between $\tilde{Q}_d(\tilde{\lambda}_1)$ and $\tilde{Q}_d(\tilde{\lambda}_2)$. It implies that a joint detection can reduce the error probability significantly. Our focus in this paper is, however, more to investigate the estimation and the modulation of nonlinear spectrum in a realistic fiber.

Simulating a lossless fiber by split step Fourier method show that the discrete spectrum can be recovered after 640km link or beyond with a high precision. The discrete spectrum of received solitons in a lossless fiber is displayed by crosses ("x") in Fig. 3. In experiment, the spectrum is distorted due to fiber loss, noise, imperfect transceivers etc. Some solitons are more sensitive to these distortions. Let us partition the constellation set into four sets: $S_1 = \{1,6,11,16\}$, $S_2 = \{2,9,8,15\}$, $S_3 = \{3,5,12,14\}$, $S_4 = \{4,7,10,13\}$ (see Fig. 3). The solitons in each partition are only different by a fixed $k\,\pi/2$ phase shift, and in principle, should be distorted similarly. We observe that $S_2$ and $S_4$ ("Λ"-shape solitons in Fig. 2) are more susceptible than $S_1$ and $S_3$ ("M"-shape solitons) to an imperfect modulation, and an imprecise launch power. Although the MZ modulator characteristic was made linear by predistorting the drive signal, Fig. 3 shows the need of higher precisions for

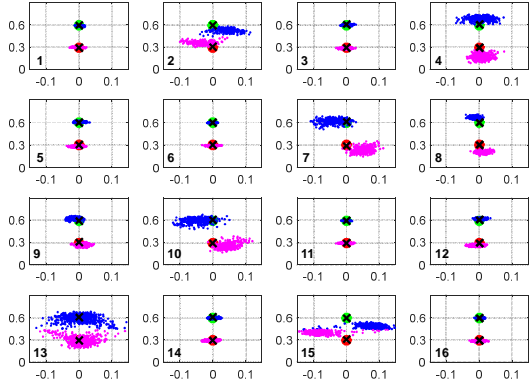
(a) $\tilde{\lambda}_1$ and $\tilde{\lambda}_2$

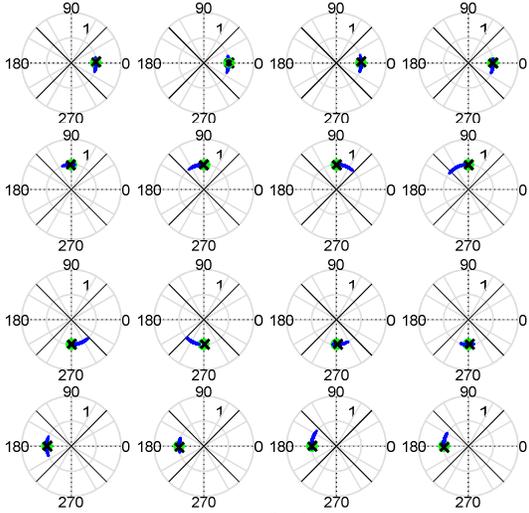
(b) $\tilde{Q}_d(\tilde{\lambda}_1)$

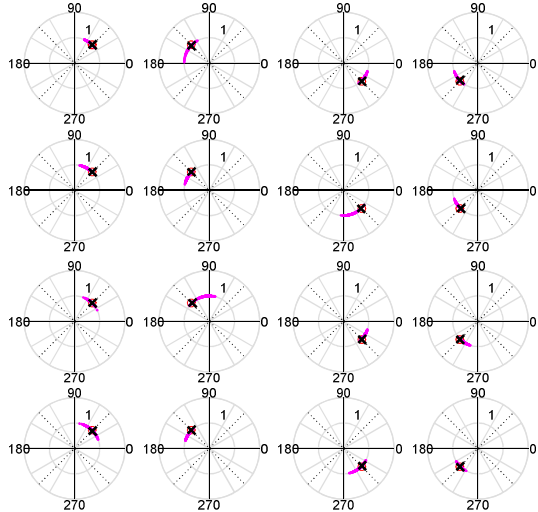
(c) $\tilde{Q}_d(\tilde{\lambda}_2)$

**Fig. 3:** Discrete spectrum of the constellation solitons, labeled from 1 to 16, after transmission over 640 km NZ-DSF Fiber (experiment) and lossless fiber (simulation). Spectrum components of the experiment (respectively, the simulation) are shown by dots (by cross "x"). $\tilde{Q}_d(\tilde{\lambda}_1)$ and $\tilde{Q}_d(\tilde{\lambda}_2)$ are transformed to the transmitter reference by Eq.(1) and normalized to 1. **(a)** 16 plots show $\tilde{\lambda}_1$ and $\tilde{\lambda}_2$ for each received soliton. **(b)** $\tilde{Q}_d(\tilde{\lambda}_1)$ for each corresponding soliton. **(c)** $\tilde{Q}_d(\tilde{\lambda}_2)$ for each corresponding soliton.

"Λ"- shape solitons. Note that "M"-shape solitons transform to "Λ"-shape solitons in a sufficiently long link and therefore, they become more sensitive in this case to small perturbations.

## Conclusions

We experimentally demonstrate 2-soliton transmission with a low error of detection. We modulate information bits over spectral phases. In principle, we can also use spectral amplitude for modulation to increase spectral efficiency. However, we observe that spectral amplitude is usually more susceptible than spectral phase to imperfect conditions, particularly, an imprecise launch power.

The solitons used in this paper are the solution of NLSE for a lossless and noiseless fiber. We observe that some of these solitons are less sensitive to realistic conditions. It is of practical interest to understand how the choice of eigenvalues, and initial discrete spectra decreases the sensitivity for a long optical link.

## References

[1] M.J. Ablowitz, et al., "The inverse scattering transform-Fourier analysis for nonlinear problems," Stud. Appl. Math. 53, 249 (1974).

[2] A. Hasegawa, T. Nyu, "Eigenvalue Communication," Journal of Lightwave Technology, vol. 11, no. 3, 1993.

[3] M.I. Yousefi, F.R. Kschischang, "Information Transmission using the Nonlinear Fourier Transform, Part I: Mathematical Tools", IEEE Trans. Inf. Theory, vol. 60, no. 7, pp. 4312–4328, Jul. 2014.

[4] M.I. Yousefi, F.R. Kschischang, "Information Transmission using the Nonlinear Fourier Transform, Part II: Numerical Methods", IEEE Trans. Inf. Theory, vol. 60, no. 7, pp. 4329–4345, Jul. 2014.

[5] M.I. Yousefi, F.R. Kschischang, "Information Transmission using the Nonlinear Fourier Transform, Part III: Spectrum Modulation", IEEE Trans. Inf. Theory, vol. 60, no. 7, pp. 4346–4369, Jul. 2014.

[6] Z. Dong, S. Hari, T. Gui, K. Zhong, M. Yousefi, C. Lu, P. Wai, F. Kschischang, A. Lau," Nonlinear Frequency Division Multiplexed Transmissions based on NFT," submitted for publication in PTL (2015).

[7] J.E. Prilepsky, S.A. Derevyanko, K.J. Blow, I. Gabitov, S.K. Turitsyn, "Nonlinear inverse synthesis and eigenvalue division multiplexing in optical fiber channels," Physical review letters, 113, 013901, (2014).

[8] S. T. Le, J. E. Prilepsky, S. K. Turitsyn," Nonlinear inverse Synthesis for High Spectral Efficiency transmission in Optical Fibers," Optics Express, Vol. 22, Issue 22, 2014.

[9] H. Bülow, "Experimental Demonstration of Optical Signal Detection Using Nonlinear Fourier Transform," J. Lightwave Technol., Vol. **33**, no. 7, pp. 1433 (2015).

[10] H. Bülow, "Nonlinear Fourier Transform Based Coherent Detection Scheme for Discrete Spectrum," Proc. OFC 2015, W3K.2, Los Angeles (2015).